\documentclass[preprint,12pt,numbers]{elsarticle}

\usepackage[utf8]{inputenc}
\usepackage{hyperref}
\usepackage{booktabs}
\usepackage{graphicx}
\usepackage{amsmath}

\usepackage{fancyhdr}

\makeatletter
\def\ps@pprintTitle{%
  \def\@oddhead{\footnotesize\itshape Submitted to Data in Brief\hfill Loth et al.}%
  \let\@evenhead\@oddhead
  \let\@oddfoot\@empty
  \let\@evenfoot\@empty
}
\makeatother
\begin{document}

\begin{frontmatter}

\title{CRED-1: An Open Multi-Signal Domain Credibility Dataset for Automated Pre-Bunking of Online Misinformation}

\author[a]{Alexander Loth\corref{cor}}
\cortext[cor]{Corresponding author.}
\ead{alexander.loth@stud.fra-uas.de}
\author[a]{Martin Kappes}
\author[b]{Marc-Oliver Pahl}
\address[a]{Frankfurt University of Applied Sciences, Frankfurt am Main, Germany}
\address[b]{IMT Atlantique, UMR IRISA, Chaire Cyber CNI, Rennes, France}

\begin{abstract}
This article presents CRED-1, an open, reproducible domain-level credibility dataset combining two openly-licensed source lists (OpenSources.co and Iffy.news) with four computed enrichment signals: domain age (WHOIS/RDAP), web popularity (Tranco Top-1M), fact-check frequency (Google Fact Check Tools API), and threat intelligence (Google Safe Browsing API). The dataset covers 2,672 domains categorized as fake, unreliable, mixed, conspiracy, or satire, each assigned a composite credibility score between 0.0 and 1.0. CRED-1 is designed for on-device deployment in privacy-preserving browser extensions to enable client-side pre-bunking of misinformation at the content delivery stage. The entire pipeline is implemented in Python using only standard library modules and is fully reproducible from publicly available sources. The dataset and pipeline code are released under CC~BY~4.0 and archived on Zenodo~\cite{loth2026cred1zenodo}.
\end{abstract}


\end{frontmatter}

\section{Specifications Table}

\noindent
\begin{tabular}{@{}ll@{}}
\toprule
\textbf{Subject} & Computer Science, Information Systems \\
\textbf{Specific subject area} & Misinformation detection, domain credibility assessment \\
\textbf{Type of data} & JSON, CSV, Python scripts \\
\textbf{Data collection} & Automated pipeline merging open source lists with API enrichment \\
\textbf{Data source location} & Online (GitHub, Zenodo) \\
\textbf{Data accessibility} & Public repository and Zenodo archive \\
\textbf{Repository name} & GitHub: \url{https://github.com/aloth/cred-1} \\
\textbf{DOI} & \href{https://doi.org/10.5281/zenodo.18769460}{10.5281/zenodo.18769460} \\
\bottomrule
\end{tabular}

\section{Value of the Data}

\begin{itemize}
\item CRED-1 provides a standardized, openly-licensed domain credibility dataset that can serve as ground truth for misinformation research and as a practical resource for content moderation tools.
\item The dataset is useful to researchers studying online misinformation, developers building browser extensions or content filters, and educators developing media literacy curricula.
\item Unlike proprietary alternatives (e.g., NewsGuard, MBFC API), CRED-1 is fully transparent, reproducible, and free to use, enabling independent verification and extension.
\item The multi-signal scoring model demonstrates how domain-level credibility can be computed from publicly available data without requiring proprietary databases or human annotation.
\item The compact JSON format (145~KB) enables on-device deployment in mobile and browser applications without server-side dependencies, preserving user privacy.
\item CRED-1 can be used as a credibility signal in automated fact-checking pipelines, complementing claim-level approaches with domain-level priors.
\item The dataset complements recent work on controlled misinformation generation~\cite{loth2026industrialized} and human credibility assessment under AI-generated content~\cite{loth2026eroding}, providing the domain-level ground truth needed for end-to-end misinformation detection pipelines.
\item Expert surveys have identified content delivery as a critical intervention point in the misinformation kill chain~\cite{loth2026verification}. CRED-1 enables automated pre-bunking at precisely this stage.
\end{itemize}

\section{Data Description}

The CRED-1 dataset is distributed in two formats:

\paragraph{Compact format} (\texttt{cred1\_v1.0.json}): A JSON object mapping domain names to credibility metadata, optimized for application embedding. Each entry contains the composite credibility score (\texttt{s}, 0.0--1.0), category code (\texttt{c}), number of independent sources (\texttt{n}), Tranco rank (\texttt{r}, optional), and domain age in years (\texttt{a}, optional). Optional fields are omitted when unavailable. A domain \emph{not present} in the dataset should be treated as neutral/unknown, not as reliable---CRED-1 is a negative-signal dataset containing only domains with known credibility issues.

\paragraph{Full format} (\texttt{cred1\_v1.0\_full.csv}): A CSV file containing all 18 fields including raw enrichment signals (Iffy.news factual rating, Iffy.news bias rating, RDAP registration date, fact-check claim count, Safe Browsing flag) and individual score components, suitable for research analysis. Rows are sorted by credibility score ascending (least credible first).

\subsection{Category Taxonomy}

Domains are classified into six categories based on consensus labels from the source datasets. When a domain appears in both sources, the lower credibility category takes precedence. Table~\ref{tab:categories} shows the distribution.

\begin{table}[h]
\centering
\begin{tabular}{@{}llrr@{}}
\toprule
\textbf{Category} & \textbf{Definition} & \textbf{Count} & \textbf{\%} \\
\midrule
Fake        & Fabricated content, deceptive, impersonation  & 493   & 18.4 \\
Conspiracy  & Consistently promotes unsupported conspiracy theories & 153   & 5.7  \\
Unreliable  & Regularly fails journalistic accuracy standards & 589   & 22.0 \\
Satire      & Humor, irony, exaggeration (not malicious)    & 94    & 3.5  \\
Mixed       & Some factual reporting alongside biased or misleading content & 1,335 & 50.0 \\
Reliable    & Generally considered reliable by fact-checkers & 8     & 0.3  \\
\midrule
\textbf{Total} & & \textbf{2,672} & \textbf{100} \\
\bottomrule
\end{tabular}
\caption{Category distribution in CRED-1. Base scores range from 0.0 (fake) to 1.0 (reliable).}
\label{tab:categories}
\end{table}

The category taxonomy is derived by mapping labels from both upstream sources into a unified scheme. OpenSources.co labels such as \emph{fake}, \emph{fake news} map to ``fake''; \emph{bias}, \emph{political}, \emph{state} map to ``mixed''; \emph{clickbait}, \emph{junksci}, \emph{hate}, \emph{rumor} map to ``unreliable.'' Iffy.news uses MBFC factual ratings: Very Low maps to ``fake,'' Low to ``unreliable,'' and Mixed to ``mixed.''

\subsection{Score Distribution}

The composite credibility score ranges from 0.000 to 0.962 with a mean of 0.299 and a standard deviation of 0.170. The distribution is bimodal, with 846 domains (31.7\%) scoring below 0.2 (primarily fake and conspiracy categories) and 1,335 domains (50.0\%) scoring between 0.4 and 0.6 (mixed-credibility domains). No domains fall in the 0.6--0.8 range, and only 8 domains (0.3\%) score above 0.8.

\subsection{Enrichment Signal Coverage}

Table~\ref{tab:coverage} summarizes the availability of each enrichment signal across the dataset.

\begin{table}[h]
\centering
\begin{tabular}{@{}lrrll@{}}
\toprule
\textbf{Signal} & \textbf{Available} & \textbf{\%} & \textbf{Range / Summary} & \textbf{Source} \\
\midrule
Source category    & 2,672 & 100.0 & 6 categories         & OpenSources + Iffy.news \\
Iffy.news score    & 2,040 & 76.3  & 0.0--1.0              & Iffy.news~\cite{iffynews2022} \\
Tranco rank        & 704   & 26.3  & 11--985,661           & Tranco~\cite{pochat2019tranco} \\
Domain age (RDAP)  & 2,325 & 87.0  & 0.9--31.6 yrs (med.\ 14.0) & Public RDAP \\
Fact-check claims  & 67    & 2.5   & 1--52 claims (332 total) & Google Fact Check~\cite{googlefactcheck} \\
Safe Browsing flag & 2     & 0.07  & Binary (flagged)      & Google Safe Browsing~\cite{googlesafebrowsing} \\
\bottomrule
\end{tabular}
\caption{Enrichment signal availability across 2,672 domains.}
\label{tab:coverage}
\end{table}

\subsection{Notable Observations}

\begin{itemize}
\item \textbf{Domain age is not a strong misinformation signal}: The median domain age of 14.0 years indicates that many misinformation sites are long-established, unlike phishing domains which tend to be recently registered. This finding contrasts with common heuristics used in cybersecurity threat detection.
\item \textbf{Most domains have zero fact-check claims}: Only 67 of 2,672 domains (2.5\%) have been specifically reviewed by fact-checkers indexed by Google's ClaimReview database. The five most fact-checked domains are: \texttt{trump.news} (52 claims), \texttt{thegatewaypundit.com} (26), \texttt{naturalnews.com} (22), \texttt{infowars.com} (8), and \texttt{breitbart.com} (7).
\item \textbf{Misinformation and malware are largely disjoint}: Only 2 of 2,672 domains were flagged by Google Safe Browsing, confirming that misinformation sites operate within the bounds of technical legitimacy while distributing misleading content.
\item \textbf{Source overlap provides validation}: 193 domains (7.2\%) appear in both OpenSources.co and Iffy.news, providing independent corroboration. The \texttt{n} (source count) field enables users to filter for higher-confidence entries.
\end{itemize}

\section{Experimental Design, Materials and Methods}

The CRED-1 pipeline consists of two phases, each implemented as a standalone Python script using only standard library modules.

\subsection{Phase~1: Source Data Acquisition and Merging}

CRED-1 aggregates domain labels from two openly-licensed source datasets:

\begin{enumerate}
\item \textbf{OpenSources.co} (CC~BY~4.0): A curated list of 825 domains classified by type (fake, bias, conspiracy, satire, unreliable, etc.), originally compiled by Zimdars~\cite{zimdars2016}. The dataset was created as part of a media literacy effort to catalog sources of misinformation and is hosted on GitHub.
\item \textbf{Iffy.news Index} (MIT license): A dataset of 2,040 domains rated as having low or very low factual reporting by Media Bias/Fact Check (MBFC), maintained by the Reynolds Journalism Institute~\cite{iffynews2022}. The index provides additional metadata including factual reporting level, political bias classification, and a numeric credibility score.
\end{enumerate}

After normalization (lowercasing, stripping \texttt{www.} prefixes, removing trailing slashes) and deduplication, the merged dataset contains 2,672 unique domains with 193 appearing in both sources. When a domain appears in both sources with conflicting category labels, the lower credibility category is assigned.

\subsection{Phase~2: Signal Enrichment}

Each domain is enriched with four independently computed signals:

\paragraph{Domain age} (WHOIS/RDAP): Registration dates are queried via the public Registration Data Access Protocol (RDAP). Successfully resolved for 2,325 domains (87\%), with a median age of 14.0 years and a range of 0.9 to 31.6 years. The remaining 347 domains returned no registration data due to RDAP server errors or missing records.

\paragraph{Web popularity} (Tranco Top-1M): The Tranco list~\cite{pochat2019tranco} provides a research-oriented aggregated popularity ranking designed to be resistant to manipulation. Matched 704 domains (26.3\%), with ranks ranging from 11 to 985,661 and 56 domains in the Top~10,000. Domains not in the Tranco list are likely low-traffic sites.

\paragraph{Fact-check frequency} (Google Fact Check Tools API~\cite{googlefactcheck}): The number of ClaimReview-annotated fact-check claims associated with each domain. Queried for all 2,672 domains; 67 (2.5\%) returned at least one claim, with 332 total claims across the dataset. Claim counts range from 1 to 52.

\paragraph{Threat intelligence} (Google Safe Browsing API~\cite{googlesafebrowsing}): Binary threat detection for malware and social engineering, checked via Google's Safe Browsing Lookup API. All 2,672 domains were checked; only 2 (0.07\%) were flagged.

\subsection{Scoring Model}

The composite credibility score $S$ is computed as a weighted blend of up to five signals:

\begin{equation}
S = w_{\text{cat}} \cdot s_{\text{cat}} + w_{\text{iffy}} \cdot s_{\text{iffy}} + w_{\text{fc}} \cdot s_{\text{fc}} + w_{\text{tranco}} \cdot s_{\text{tranco}} + w_{\text{age}} \cdot s_{\text{age}} + w_{\text{fill}} \cdot s_{\text{cat}}
\label{eq:score}
\end{equation}

where the base weights are $w_{\text{cat}} = 0.50$, $w_{\text{iffy}} = 0.15$, $w_{\text{fc}} = 0.15$, $w_{\text{tranco}} = 0.05$, $w_{\text{age}} = 0.05$, and $w_{\text{fill}}$ compensates for missing signals by reverting their weight to the category score. When all signals are available, $w_{\text{fill}} = 0$. Individual signal scores are computed as follows:

\begin{itemize}
\item $s_{\text{cat}}$: Category lookup (fake $= 0.0$, conspiracy $= 0.1$, unreliable $= 0.2$, satire $= 0.3$, mixed $= 0.5$, reliable $= 1.0$).
\item $s_{\text{iffy}}$: Raw Iffy.news credibility score (already normalized to 0.0--1.0).
\item $s_{\text{fc}}$: $\max(0, 1 - \log_{10}(\text{claims}) / 1.7)$, yielding 0.82 for 1 claim and approaching 0.0 for 50+ claims.
\item $s_{\text{tranco}}$: $\max(0, 1 - \log_{10}(\text{rank}) / 6)$, mapping rank~1 to 1.0 and rank~1{,}000{,}000 to 0.0.
\item $s_{\text{age}}$: $\min(1, \text{age\_years} / 20)$, saturating at 20 years.
\end{itemize}

\textbf{Override}: Domains flagged by Google Safe Browsing receive a hard score cap of $S = 0.05$, regardless of other signals.

\subsection{Reproducibility}

The entire pipeline is implemented in two Python scripts (\texttt{build\_dataset.py} and \texttt{enrich\_dataset.py}) using only standard library modules (\texttt{json}, \texttt{csv}, \texttt{urllib}, \texttt{zipfile}). No external packages are required. The Google Fact Check Tools API and Safe Browsing API require a free API key from Google Cloud Console; the key can be provided via the \texttt{GOOGLE\_API\_KEY} environment variable or macOS Keychain. Complete reproduction from source data to final dataset takes approximately 30~minutes on a standard internet connection. SHA-256 checksums are provided in the repository's \texttt{CODEBOOK.md} for integrity verification.

\subsection{Limitations}

\begin{itemize}
\item \textbf{English-language bias}: The majority of domains in both upstream sources are English-language outlets. Coverage of non-English misinformation sources is limited.
\item \textbf{Temporal validity}: Domain credibility can change over time. CRED-1 v1.0 reflects the state of source data as of February 2026. Periodic updates are planned.
\item \textbf{Negative-signal design}: CRED-1 contains only domains with known credibility issues. Absence from the dataset does not indicate reliability.
\item \textbf{Upstream dependency}: The dataset inherits any biases or errors present in the OpenSources.co and Iffy.news source lists.
\end{itemize}

\section*{Ethics Statement}

CRED-1 aggregates publicly available domain-level metadata and does not contain personal data. All source datasets are openly licensed (OpenSources.co under CC~BY~4.0; Iffy.news under MIT). The Google Fact Check Tools API and Safe Browsing API were used in accordance with their terms of service. The dataset reflects credibility assessments by independent fact-checking organizations and should be interpreted as one signal among many in any decision-making context.

\section*{CRediT Author Statement}

\textbf{Alexander Loth}: Conceptualization, Methodology, Software, Data curation, Writing -- original draft, Writing -- review \& editing. \textbf{Martin Kappes}: Supervision. \textbf{Marc-Oliver Pahl}: Supervision.

\section*{Declaration of Competing Interest}

The author declares no competing interests.

\section*{Data Availability}

The dataset and reproduction pipeline are available at \url{https://github.com/aloth/cred-1} and permanently archived on Zenodo at \href{https://doi.org/10.5281/zenodo.18769460}{10.5281/zenodo.18769460}~\cite{loth2026cred1zenodo}.

\section*{Acknowledgments}

The author thanks Melissa Zimdars and the OpenSources.co project for their pioneering work in cataloging unreliable news sources. The author is grateful to the Iffy.news team at the Reynolds Journalism Institute for maintaining the Iffy Index. This work uses the Google Fact Check Tools API and Google Safe Browsing API, provided free of charge by Google.

\bibliographystyle{elsarticle-num}

\end{document}